\documentclass[aps,prd,reprint,amsmath,amssymb]{revtex4-2}
\usepackage[T1]{fontenc}
\usepackage[utf8]{inputenc}
\usepackage{babel}
\usepackage[pdfusetitle,
 bookmarks=true,bookmarksnumbered=false,bookmarksopen=false,
 breaklinks=false,pdfborder={0 0 1},backref=false,colorlinks=false]
 {hyperref}
\usepackage{graphicx}
\usepackage{float}
\usepackage{ulem}
\usepackage{xcolor}
\usepackage[cal=boondox]{mathalfa}

\renewcommand{\eqref}[1]{Eq.~(\ref{#1})}
\newcommand{\figref}[1]{Fig.~\ref{#1}}

\begin{document}

\title{First Order Phase Transition Induced Graviton Bremsstrahlung: A Multi Peak Gravitational Wave Signature}

\author{Dipendu Bhandari}
\email{dbhandari@iitg.ac.in}

\author{Rajat Kumar Mandal}
\email{r.mandal@iitg.ac.in}

\author{Arunansu Sil}
\email{asil@iitg.ac.in}

\affiliation{Department of Physics, Indian Institute of Technology Guwahati, Assam
781039, India.}

\begin{abstract}
Gravitational waves (GWs) from first order phase transitions (FOPTs) are conventionally sourced by bubble collisions, sound waves, and plasma turbulence. 
We propose a novel microscopic GW source arising from graviton bremsstrahlung during the decay of the scalar field driving the FOPT. 
In the presence of Yukawa interactions with light fermions, the scalar decay is inevitably accompanied by graviton emission due to the universal coupling of gravity to the energy-momentum tensor.
We show that these gravitons generate an additional stochastic GW background during the phase-transition epoch, complementing the conventional FOPT signal. 
The resulting GW spectrum can exhibit a characteristic multi-peaked structure. 
Unlike scenarios in which different GW components originate from distinct cosmological epochs, all contributions considered here emerge from the dynamics of the same FOPT, offering a unique probe of both its microscopic particle dynamics and macroscopic plasma evolution.
\end{abstract}

\maketitle

\section{Introduction}

Gravitational waves (GW)~\cite{LIGOScientific:2016aoc} provide a unique observational window into the early Universe. 
Due to their extremely weak interaction with matter, GWs propagate almost unimpeded from their production epoch, carrying information about physical processes that generate them at different stages of the early Universe such as inflation, reheating, cosmological phase transitions~\cite{Mazumdar:2018dfl} till a much later stage where they can also be generated by astrophysical events such as black hole and neutron star mergers \cite{LIGOScientific:2016aoc,LIGOScientific:2017vwq}. Among these, strongly first-order phase transitions (FOPT) are especially compelling as their out-of-equilibrium dynamics can source an observable stochastic GW background.
While the electroweak phase transition in the Standard Model (SM) is expected to be a smooth crossover \cite{Kajantie:1995kf,Kajantie:1996mn,DOnofrio:2015gop}, many beyond the SM (BSM) scenarios accommodate strongly FOPTs~\cite{Mazumdar:2018dfl} capable of generating detectable GWs.
Furthermore, FOPTs play a central role in several baryogenesis scenarios \cite{Morrissey:2012db,Cohen:1993nk,Trodden:1998ym} by providing the necessary departure from thermal equilibrium.
Consequently, the observation of a stochastic GW background originating from an FOPT would not only reveal crucial information about the thermal evolution of the early Universe but also provide compelling evidence for physics beyond the SM.

During an FOPT, the Universe evolves from a metastable (symmetric phase) to a stable (broken phase) state via the nucleation and expansion of true-vacuum bubbles.
GWs are produced when these bubbles expand, collide, and merge, generating anisotropic stresses \cite{Kosowsky:1992rz,Kosowsky:1992vn}.
Additional GW sources arise when expanding bubble walls interact with the primordial plasma, converting part of the released vacuum energy into bulk fluid motion.
The resulting long lived sound waves, sustaining beyond the duration of phase transition, can efficiently source GWs \cite{Hindmarsh:2013xza,Hindmarsh:2015qta}, often dominating over the bubble-collision signal.
Subsequent shocks and turbulence in the plasma may further enhance the stochastic GW background \cite{Kosowsky:2001xp,Caprini:2006jb} at a later stage.

\begin{figure}[h]
\centering
\includegraphics[width=0.8\linewidth]{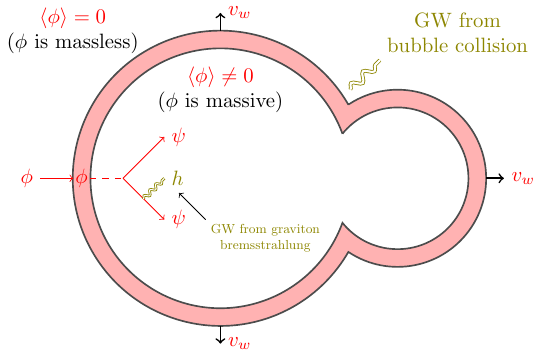}
\caption{Schematic diagram of emission of gravitons during bubble nucleation.}
\label{fig:Schematic diagram of GW emission-1}
\end{figure}

Here we propose a novel source of gravitational waves generated during the FOPT, beyond the above mentioned conventional macroscopic GW sources associated with bubble collisions and the subsequent acoustic and turbulent dynamics of the plasma. 
As these bubbles nucleate and expand, regions of the symmetric phase are converted into the broken phase where the scalar field $\phi$ responsible for FOPT acquires a nonzero vacuum expectation value (\textit{vev}). 
Consequently, $\phi$ particles inside the bubbles become massive and in presence of a Yukawa interaction of the form $y_{\phi}\phi\bar{\psi}\psi$, they can decay via the process $\phi \rightarrow \bar{\psi}+\psi$ where 
$\psi$ is representative of a BSM fermion. 
Upon minimal coupling to gravity \cite{Donoghue:1994dn,Holstein:2006bh}, the universal interaction of the graviton with the energy-momentum tensor inevitably induces the accompanying three-body decay channel $\phi \rightarrow \bar{\psi}+\psi+{h}$, where $h$ denotes a graviton \cite{Holstein:2006bh}. 
The gravitons emitted through such \textit{graviton bremsstrahlung} \cite{Weinberg:1965nx,Nakayama:2018ptw,Huang:2019lgd,Ghoshal:2022kqp,Barman:2023ymn,Bernal:2023wus,Bernal:2023wus,Hu:2024bha,Strumia:2025dfn,Datta:2024tne,Datta:2025wfh} process generate an additional stochastic GW background during the phase transition epoch itself, providing a previously unexplored correlated microscopic source of gravitational radiation during FOPT, schematically illustrated in \figref{fig:Schematic diagram of GW emission-1}. 
This contribution complements the standard GW signals generated by the macroscopic dynamics of the FOPT and gives rise to a characteristic multi-peaked GW spectrum. 

Multi-peaked GW spectra have been discussed previously in several contexts~\cite{Borah:2026zbl,Borah:2024lml,Borboruah:2022eex,Ai:2025fqw}. 
In multi-step phase transitions, distinct peaks can arise \cite{Benincasa:2022elt,Morais:2019fnm} from successive symmetry-breaking transitions occurring at different stages of the thermal history. 
Alternatively, in scenarios involving domain walls \cite{Saikawa:2017hiv,Hiramatsu:2013qaa}, cosmic strings~\cite{Jana:2025vyb} or primordial black holes \cite{Lewicki:2023ioy,Kodama:1982sf} produced during a first-order phase transition, the conventional GW signal from bubble collisions and acoustic plasma effects can be supplemented by additional GW components generated through the subsequent annihilation of domain walls or the later dynamics of primordial black holes. 
In such cases, the different GW peaks either originate from multiple phase transitions or are associated with distinct cosmological events occurring after the completion of the FOPT. 
Moreover, the domain-wall and primordial-black-hole scenarios typically require additional ingredients beyond the FOPT itself, such as extra discrete symmetries giving rise to topological defects \cite{Saikawa:2017hiv}.

In contrast, the mechanism proposed here neither relies on multiple phase transitions nor invokes secondary cosmological relics. 
The additional GW component originates directly from graviton bremsstrahlung accompanying the decay of the same scalar field responsible for the FOPT and is generated during the phase-transition epoch itself. 
Since the existence of this bremsstrahlung channel follows solely from the universal coupling of gravity to the energy-momentum tensor, the mechanism is generic and can be formulated in a largely model-independent manner. Consequently, unlike existing multi-peaked GW scenarios where different peaks are traced back to distinct cosmological events, all peaks in our framework originate from different manifestations of the same phase transition. 
To clarify further and for a comprehensive discussion, we present below first the origins of conventional GW emissions during an FOPT in brief and, then plan to discuss in detail the \textit{graviton bremsstrahlung} process during the FOPT and estimate the resulting GW spectrum thereafter. 

\section{Conventional Gravitational Waves from FOPT}
\begin{figure}[h]
    \centering
    \includegraphics[width=1\linewidth]{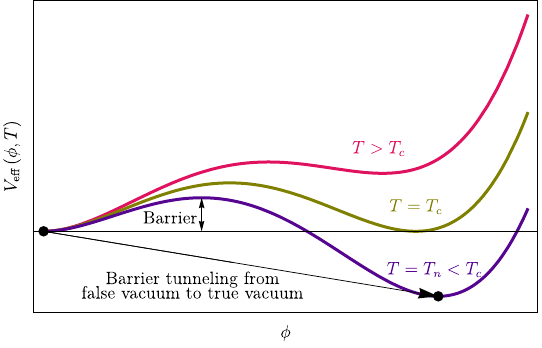}
    \caption{Schematic diagram of an FOPT.}
    \label{fig:FOPT-diagram}
\end{figure}

A first-order phase transition~\cite{Mazumdar:2018dfl} occurs when the finite-temperature effective potential develops two degenerate minima separated by a potential barrier at the critical temperature $T_c$. As the Universe cools further, the system transits from the symmetric phase ($\langle \phi \rangle =0$) to the broken phase ($\langle\phi \rangle \neq 0$) via quantum tunneling. This is shown in the Fig.~\ref{fig:FOPT-diagram} schematically.
This tunnelling process leads to the nucleation of bubbles of the broken phase (inside which the $\phi$ \textit{vev} is non-zero), which subsequently expand within the surrounding plasma and eventually collide with each other, thereby converting the whole Universe into the broken phase.

The FOPT in the early Universe provides a well-motivated source of stochastic GWs~\cite{Kosowsky:1992rz,Kosowsky:1992vn,Hindmarsh:2013xza,Hindmarsh:2015qta,Kosowsky:2001xp,Caprini:2006jb}, arising from the macroscopic dynamics of bubble nucleation in the plasma. 
Once nucleated, the expanding bubbles and their interaction with the surrounding plasma give rise to three primary sources of GW production: the bubble collisions, the sound waves in the plasma, and magnetohydrodynamic (MHD) turbulence as we discuss below. 

\subsection{Bubble Collisions}

When the bubbles collide during nucleation, the spherical symmetry breaks which sources the anisotropy in the stress-energy tensor, leading to the generation of GW. 
Using envelop approximation, the GW spectrum in this case can be evaluated as~\cite{Jinno:2016vai,Athron:2023xlk}
\begin{widetext}
\begin{equation}
    \mathcal{h}^2 \Omega_{\rm GW}^{\rm col} = \underbrace{ 1.67 \times 10^{-5} \left(\frac{g_{*s}}{100}\right)^{-\frac{1}{3}}}_{\text{Redshift}} \overbrace{\left( \frac{\beta}{H}\right)^{-2} \left( \frac{k_{\rm col} \alpha}{1 + \alpha} \right)^2 \Delta(v_w)}^{\text{\rm Scaling}} \, \underbrace{S_{\rm col} (f)}_{\text{Spectral Shape}}.
\end{equation}
\end{widetext}
In the above expression, the first part, labeled as \textit{Redshift}, accounts for the cosmological dilution of the GW energy density from the time of production to the present epoch, where $g_{*s}$ denotes the effective number of degrees of freedom. The second expression, denoted as \textit{Scaling}, encodes the dynamics of the phase transition. In particular, the factor $(\beta/H)^{-2}$ reflects the duration of the transition, with smaller $\beta$ corresponding to a longer-lasting transition and hence a larger GW amplitude. The parameter $\alpha$ characterizes the strength of the transition and is defined as the ratio of the released vacuum energy (latent heat) to the radiation energy density of the Universe, while $k_{\rm col}$ represents the efficiency with which the released vacuum energy is converted into scalar field gradient energy. The function $\Delta(v_w)$ captures the dependence on the bubble wall velocity.
Finally, $S_{\rm col}(f)$~\cite{Jinno:2016vai} describes the spectral shape of the GW signal, which determines how the energy density is distributed over frequency. It is normalized such that $S_{\rm col}(f_{\rm col}^{\rm peak}) = 1$, with $f_{\rm col}^{\rm peak}$ denoting the peak frequency of the spectrum.

\subsection{Sound Waves}
In FOPT, the expanding bubble walls transmit a significant fraction of the released vacuum energy to the surrounding plasma. As a result, during and after the bubble collisions, the kinetic energy fluctuations of the plasma leads to the acoustic wave generation. 
These sound waves correspond to the relativistic bulk motion of the plasma carrying energy which enters into the energy-momentum tensor and can directly modify the spacetime curvature, inducing the generation of GWs.

The GW signal amplitude from sound waves determined from the hydrodynamic simulations~\cite{Hindmarsh:2013xza,Hindmarsh:2015qta,Athron:2023xlk} can be expressed as
\begin{widetext}
\begin{equation}
    \mathcal{h}^2 \Omega_{\rm GW}^{\rm sw} = \underbrace{ 2.59 \times 10^{-6} \left(\frac{g_{*s}}{100}\right)^{-\frac{1}{3}}}_{\text{Redshift}} \overbrace{\left( \frac{\beta}{H}\right)^{-1} \left( \frac{k_{\rm sw} \alpha}{1 + \alpha} \right)^2 v_w \Upsilon(\tau_{\rm sw})}^{\text{\rm Scaling}} \, \underbrace{S_{\rm sw} (f)}_{\text{Spectral Shape}}.
\end{equation}
\end{widetext}
Following the discussion of the bubble collision contribution, the sound wave signal can similarly be understood in terms of redshift and scaling structures, with key differences arising from the hydrodynamic nature of the source. In particular, while the dependence on the transition strength $\alpha$ and duration $\beta$ remains as discussed previously, the dominant energy transfer now occurs into the bulk motion of the plasma.
The efficiency factor $k_{\rm sw}$~\cite{Espinosa:2010hh} quantifies the fraction of vacuum energy converted into fluid motion, replacing the scalar-field dominated contribution relevant for bubble collisions. The bubble wall velocity $v_w$ plays a more direct role in this case, as it governs how effectively the expanding walls accelerate the plasma and sustain the acoustic waves. Additionally, the factor $\Upsilon(\tau_{\rm sw})$ accounts for the finite lifetime of the sound wave source~\cite{Guo:2020grp}, leading to a suppression of the GW signal when the acoustic source becomes short-lived compared to a Hubble time. The lifetime of the source $\tau_{\rm sw}$ is defined as 
\begin{align}
    \tau_{\rm sw} \sim \frac{R_*}{\overline{U}_f},
\end{align}
where, $R_* = (8\pi)^{\frac{1}{3}} v_w \beta^{-1}$ is the mean bubble separation and $\overline{U}_f = \sqrt{\frac{3 k_{\rm sw} \alpha}{4 (1+ \alpha)}}$ denotes the RMS fluid velocity.
The term $S_{\rm sw}(f)$~\cite{Caprini:2015zlo} again represents the spectral shape of the GW signal and is normalized such that $S_{\rm sw}(f^{\rm peak}_{\rm sw}) = 1$, where $f_{\rm sw}^{\rm peak}$ denotes the peak frequency of the sound wave contribution.

\subsection{Magnetohydrodynamic Turbulence}
Magnetohydrodynamic (MHD) turbulence can develop in the plasma following bubble expansion and collisions during a first-order phase transition, as large amounts of energy are injected into the fluid at characteristic length scales set by the bubble size. The resulting bulk motions can generate shocks, which subsequently evolve into turbulent flows characterized by irregular eddies. 
In this regime, energy cascades from large eddies (of order the bubble radius) to smaller scales, sustaining stochastic fluid motion. These turbulent motions contribute to the energy-momentum tensor of the plasma and act as another source of gravitational waves, producing a stochastic GW background with a characteristic spectrum determined by the transition dynamics.

The GW contribution from MHD turbulence~\cite{Kosowsky:2001xp,Caprini:2006jb,Athron:2023xlk} can be written as
\begin{widetext}
\begin{equation}
    \mathcal{h}^2 \Omega_{\rm GW}^{\rm turb} = \underbrace{ 3.34 \times 10^{-4} \left(\frac{g_{*s}}{100}\right)^{-\frac{1}{3}}}_{\text{Redshift}} \overbrace{\left( \frac{\beta}{H}\right)^{-1} \left( \frac{k_{\rm turb} \alpha}{1 + \alpha} \right)^{\frac{3}{2}} v_w}^{\text{\rm Scaling}} \, \underbrace{S_{\rm turb} (f)}_{\text{Spectral Shape}}.
\end{equation}
\end{widetext}

The efficiency factor $k_{\rm turb}$ characterizes the fraction of the released vacuum energy that is transferred into turbulent plasma motion. Compared to the sound wave case, the scaling with $(k_{\rm turb}\alpha)^{3/2}$ reflects the stochastic and non-linear nature of turbulence. The bubble wall velocity $v_w$ again controls the efficiency of energy injection into the plasma, thereby influencing the strength of the turbulent source.
Finally, the spectral shape function $S_{\rm turb}(f)$~\cite{Caprini:2015zlo} encodes the frequency distribution of the GW signal sourced by MHD turbulence, which typically exhibits a broader spectrum due to the cascade of energy across multiple length scales in the plasma.

\section{Graviton Bremsstrahlung during FOPT}

Having reviewed the conventional GW sources associated with a FOPT, we now turn to our proposal for the additional stochastic GW background arising from graviton bremsstrahlung accompanying the decay of the scalar field $\phi$ during the same FOPT. 
In materializing the above proposal, we first consider a FOPT driven by the scalar field $\phi$ in the early Universe at a temperature $T$ through the nucleation of bubbles, inside which the \textit{vev} of $\phi$ satisfies $v_\phi \neq 0$, in the background of the symmetric phase where $v_\phi = 0$. Outside the bubble, $\phi$ field remains massless and therefore it is in thermal equilibrium with the SM plasma. 

During bubble nucleation, as soon as a region of the symmetric phase is engulfed by the expanding bubble, the $\phi$ particles therein become massive (with mass $m_{\phi}$) instantaneously due to the discontinuous change of the $\phi$ \textit{vev} from zero (outside the bubble) to a nonzero value (inside). 
At this point, in presence of the Yukawa interaction Lagrangian given by 
\begin{align}
    - \mathcal{L}_{\rm int} \supset y_\phi \phi \bar{\psi} \psi , \label{eq:Yukawa interaction phi psi psi}
\end{align}
where $y_\phi$ represents the coupling strength between the scalar field $\phi$ and the BSM fermion $\psi$ of mass $m_{\psi}$, the field $\phi$ starts to decay out-of-equilibrium (provided $m_\phi > T$ and $m_{\psi} \ll m_\phi$) via $ \phi \rightarrow \bar{\psi} \psi$. 

Interestingly, besides this two-body decay channel of $\phi$, the three-body decay channel $\phi \rightarrow \bar{\psi} \psi {h}$ becomes inevitable providing additional source of the graviton ($h$) emission during bubble expansion. 
The origin of these graviton emission is based upon the action functional, given by
\begin{align}
    S\left[g,\phi,\psi\right] \supset \int d^{4}x\sqrt{-g}\left[\frac{2}{k^2} \mathcal{R}\left(g\right)+\mathcal{L}_\phi^{\rm kin} + \mathcal{L}_\psi^{\rm kin} + \mathcal{L}_{\rm int} \right],  \label{eq:Action functional}
\end{align}
where, $\mathcal{R}$ denotes the Ricci scalar and $k=2/M_p$ with $M_p = 2.4 \times 10^{18}$ GeV.
\begin{figure*}
    \includegraphics[width=0.24\textwidth]{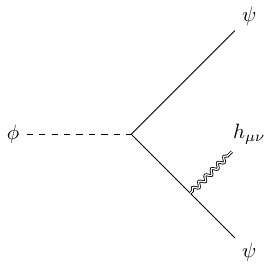} \includegraphics[width=0.24\textwidth]{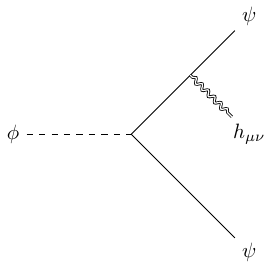} \includegraphics[width=0.24\textwidth]{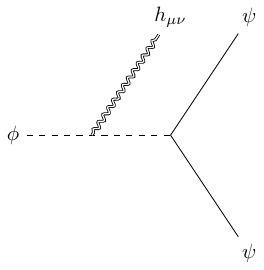} \includegraphics[width=0.24\textwidth]{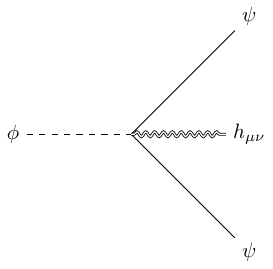}
    \caption{Feynman diagrams contributing to the graviton production.}
    \label{fig:feynman-diagrams}
\end{figure*}

In the weak field approximation, the metric $g_{\mu\nu}$ can be perturbed as $g_{\mu\nu} = \eta_{\mu\nu} + k {h}_{\mu\nu}$, with $\eta_{\mu\nu}$ denoting the flat Minkowski metric. 
Subsequently, couplings of canonically normalized graviton ${h}_{\mu\nu}$ with the stress-energy tensors $T^{\mu\nu}_\phi$ and $T^{\mu\nu}_\psi$ will appear due to the kinetic terms in the action functional in Eq.~\ref{eq:Action functional}. 
This interaction Lagrangian of graviton at the first order in $k$ is expressed as~\cite{Choi:1994ax,Holstein:2006bh} 
\begin{align}
 - \mathcal{L}_{\rm int}^h \supset \frac{k}{2} h_{\mu\nu} \sum_{X= \phi, \psi} T^{\mu\nu}_X , \label{eq:Graviton interaction Lagrangian}
\end{align}
where, the stress-energy tensors in the flat background for the scalar field $\phi$ and the fermion $\psi$ are given by 
\begin{eqnarray}
    T^{\mu\nu}_\phi = \partial^\mu \phi \partial^\nu \phi - \eta^{\mu\nu} \left( \frac{1}{2} \partial^\alpha \phi \partial_\alpha \phi - V(\phi) \right) , \nonumber \\ 
    T^{\mu\nu}_\psi = \frac{i}{4} \left( \bar{\psi} \gamma^\mu \partial^\nu \psi + \bar{\psi} \gamma^\nu \partial^\mu \psi  \right) -\eta^{\mu\nu} \nonumber \\
    \times \left( \frac{i}{2} \bar{\psi} \gamma^\alpha \partial_\alpha \psi - m_\psi \bar{\psi} \psi \right). \label{eq:Stress-energy Tesnors of phi and psi}
\end{eqnarray}

The framework now accommodates the three-body decay processes $\phi \rightarrow \bar{\psi} \psi h $, as evident from Eqs.~(\ref{eq:Yukawa interaction phi psi psi}) and (\ref{eq:Graviton interaction Lagrangian}), where graviton is produced via bremsstrahlung mechanism during the FOPT. 
The relevant Feynman rules for $h-\phi$ and $h-\psi$ couplings are taken from \citep{Holstein:2006bh} while the corresponding Feynman diagrams giving rise to graviton production from $\phi$ decay are given in Fig.~\ref{fig:feynman-diagrams}. The last two diagrams of \figref{fig:feynman-diagrams} are zero because $T^{\mu\nu}_{\phi}\epsilon_{\mu\nu}=0$ and $\eta^{\mu\nu}\epsilon_{\mu\nu}=0$. Since the GW emission via this \textit{graviton bremsstrahlung} is intricately connected to the details of the FOPT and related parameters, we plan to devote the following subsection on dynamics of the FOPT, and thereafter we proceed for the estimation of the GW spectrum.

\subsection{Dynamics of FOPT}
In this section, we summarize the FOPT dynamics with some key equations in a model-independent way. 
In FOPT, the Universe tunnels from the false vacuum state to the true vacuum state. 
This tunneling rate per unit volume is known as the false vacuum decay rate and has an approximate exponential form~\cite{Coleman:1977py,Enqvist:1991xw} 
\begin{align}
    \Gamma_d(t)= \Gamma_0 e^{\beta t}, \label{eq:False vacuum decay rate-exponential approx}
\end{align}
with the assumption $\beta/H \gg 1$. 
Here, $t$ denotes the time and $\beta^{-1}$ represents the timescale of the duration of the phase transition. The fraction of the spatial volume of the Universe remaining in the false vacuum at time $t$ is given by~\cite{Turner:1992tz}
\begin{align}
    P_{\rm s}(t) = e^{-I(t)} , \label{eq:False vacuum fraction-1}
\end{align}
where, the quantity $I(t)$ is expressed as
\begin{align}
    I(t) = \frac{4\pi}{3} \int_{t_c}^t dt' \Gamma_d(t') a^3(t') R(t;t') . \label{eq:I(t)}
\end{align}
Here, $a(t)$ is the scale factor, $t_c$ represents the time corresponding to the critical temperature $T_c$ and $R(t;t')$, denoting the comoving radius of the true vacuum bubble at time $t$ after forming at some earlier time $t'$, can be defined as
\begin{align}
    R(t;t') = \int_{t'}^t d\tilde{t} \frac{v_w}{a(\tilde{t})} , \label{eq:bubble comoving radius}
\end{align}
where, $v_w$ is the bubble wall velocity.

Assuming that the phase transition completes within one Hubble time, \textit{i.e.} $\beta/H \gg 1$, the scale factor $a(t)$ can be considered constant during the transition and $P_{\rm s}(t)$ can be analytically expressed as
\begin{align}
    P_{\rm s}(t) = {\rm exp} \left[ - I_0 e^{-\beta t}  \right] , \label{eq:False vacuum fraction-2}
\end{align}
where the factor $I_0$ is given by $I_0 = {8\pi v_w^3}\Gamma_0/{\beta^4}$. At $t= t_p$, known as the percolation time, $P_{\rm s}$ satisfies $P_{\rm s}(t_p) = {\rm exp} \left[ - I_0 e^{-\beta t_p}  \right] = 0.71$. 
From this relation, $\Gamma_0$ can be calculated and after substituting this into the \eqref{eq:False vacuum fraction-2}, we get 
\begin{align}
    P_{\rm s}(t) = {\rm exp} \left[ - \ln(0.71)\, e^{-\beta (t-t_p)}\right]. 
    \label{eq:False vacuum fraction-3}
\end{align}
Using the time-temperature adiabatic relation 
\begin{align}
    \frac{dT}{dt} = -H(T) T ,
\end{align}
the fraction of the spatial volume of the Universe remaining in the false vacuum state at temperature $T$, $P_{\rm s}(T)$ can now be written as
\begin{align}
    {P}_{\rm s}(T) = {\rm exp} \left[ \ln(0.71) \, \left(\frac{T}{T_p}\right)^{-\beta/H_p} \right] , \label{eq:False vacuum fraction-4-Temperature dependence}
\end{align}
where, $T_p$ demotes the percolation temperature and $H_p$ is the Hubble expansion rate at $T=T_p$.
\begin{figure}[H]
	\centering
	\includegraphics[width=\linewidth]{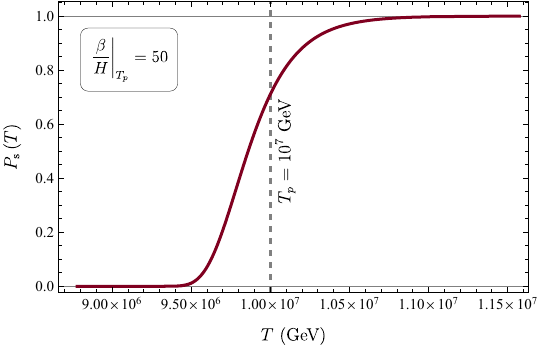}
	\caption{Progress of the FOPT as the Universe cools down.}
    \label{fig:False vacuum fraction}
\end{figure}

When the  probability of forming at least one bubble per horizon volume reaches $\mathcal{O}(1)$, the transition from the false to the true vacuum effectively begins, initiating bubble nucleation. 
The characteristic temperature at which this occurs is known as the nucleation temperature $T_n$ that can be calculated using the relation
\begin{align}
    \Gamma_d(T_n) \simeq H(T_n) . \label{eq:Nucleation temperature relation}
\end{align}
The variation of the quantity false-vacuum fraction $P_{\rm s}(T)$ (and the corresponding nucleation temperature $T_n$) is shown in the Fig.~\ref{fig:False vacuum fraction} for a set of benchmark parameters. 
This quantity is of utmost importance in our analysis since the production of gravitons via \textit{graviton bremsstrahlung} would sustain from $T_n$ until the completion of the phase transition at temperature $T_f$, beyond which the $\phi$ number density
becomes negligible and its decay no longer contributes to graviton
production. Below we proceed to estimate the corresponding GW spectrum during the FOPT. 

\subsection{Estimating GW spectrum from graviton bremsstrahlung}

After the true-vacuum bubble nucleation starts, the massless $\phi$ particles getting inside the bubble become massive and start to decay out-of-equilibrium through $\phi \rightarrow \bar{\psi} \psi$ with decay width given by 
\begin{align}
    \Gamma_{\phi\rightarrow\bar{\psi}\psi}=\frac{y^{2}_{\phi}}{8\pi}m_{\phi}\left(1-\frac{4m^{2}_{\psi}}{m^{2}_{\phi}}\right)^{\frac{3}{2}}, \label{eq:2-body-decay-rate}
\end{align}
while the three-body differential decay width for the process $\phi \rightarrow \bar{\psi} \psi h$ can be expressed as \cite{Nakayama:2018ptw}
\begin{widetext}
\begin{eqnarray}
\frac{d\Gamma_{\phi\rightarrow\bar{\psi}\psi h}}{dk_{e}}=\frac{y^{2}_{\phi}}{64\pi^{3}}\left(\frac{m_{\phi}}{M_{\text{p}}}\right)^{2}\left[\frac{1}{x}\alpha\left(1-2x\right)\left(8xy^{2}+2x\left(x-1\right)-8y^{4}-2y^{2}+1\right) \right. \nonumber \\
 \left. +4y^{2}\left(y^{2}\left(5-8x\right)-\left(x-1\right)^{2}-4y^{4}\right)\ln{\left(\frac{1+\alpha}{1-\alpha}\right)}\right], \label{eq:3-body-ddr}
\end{eqnarray}
\end{widetext}
where $\alpha=\sqrt{1-\frac{4y^{2}}{1-2x}}$, $x=\frac{k_{e}}{m_{\phi}}$
and $y=\frac{m_{\psi}}{m_{\phi}}$. Here, $k_{e}$ is the graviton momentum
(equivalent to the energy $E_e$ since gravitons are massless) at the time of emission.

During the expansion of broken-phase bubbles, let $dN_{\phi}$ denote the number of $\phi$ particles that enter the bubble interior between times $t$ and $t+dt$. Then the quantity
\begin{equation}
\frac{dN_{\phi}}{dt}
\frac{d\Gamma_{\phi\rightarrow\bar{\psi}\psi h}}
{\Gamma_{\phi\rightarrow\bar{\psi}\psi}},
\end{equation}
represents the differential production rate of gravitons, \textit{i.e.,} the number of gravitons emitted per unit time through the three-body decay $\phi\rightarrow\bar{\psi}\psi h$ with energies in the interval $[E_e,E_e+dE_e]$. Here,
$\frac{d\Gamma_{\phi\rightarrow\bar{\psi}\psi h}}
{\Gamma_{\phi\rightarrow\bar{\psi}\psi}}$
is the differential branching fraction for graviton emission into the corresponding energy bin.
The energy carried by these gravitons is
\begin{equation}
\frac{dN_{\phi}}{dt} E_e
\frac{ d\Gamma_{\phi\rightarrow\bar{\psi}\psi h}}
{\Gamma_{\phi\rightarrow\bar{\psi}\psi}},
\end{equation}
which corresponds to the GW energy emitted per unit time in the energy interval $[E_e,E_e+dE_e]$.
Here we note that even though in the IR limit $x\rightarrow 0$ the differential decay rate \eqref{eq:3-body-ddr} is singular, the total energy released $E_e\frac{d\Gamma_{\phi\rightarrow\bar{\psi}\psi h}}{dk_{e}}$ is IR safe \cite{Weinberg:1965nx}.

To obtain the average GW energy injection rate, we coarse-grain over spatial volume containing many bubbles. The corresponding GW energy density production rate in the energy interval $\left[E_e, E_e+dE_e \right]$ is given by
\begin{align}
& \frac{1}{\mathcal{V}_{\text{tot}}}  \frac{n^{\text{eq}}_{\phi} d\mathcal{V}_{\text{B}}}{dt} E_e \frac{d\Gamma_{\phi\rightarrow\bar{\psi}\psi h}}{\Gamma_{\phi\rightarrow\bar{\psi}\psi}} \nonumber \\
=& - H T \, n^{\text{eq}}_{\phi} \,\frac{d P_{\rm s}(T)}{dT} \, E_{e} \, \frac{d\Gamma_{\phi\rightarrow\bar{\psi}\psi h}}{\Gamma_{\phi\rightarrow\bar{\psi}\psi}},  \label{eq:GW energy density /time}
\end{align}
where $\mathcal{V}_{\rm tot}$ denote the total volume of the Universe.
For an infinitesimal broken-phase volume element $d\mathcal{V}_{\rm B}$ swept out by the expanding bubble wall, we assume that the $\phi$
particles initially retain the equilibrium number density of the
symmetric phase. Since the $\phi$ particles acquire a large mass upon
entering the broken phase, they decay promptly before their number
density can be significantly modified by subsequent interactions.
Therefore, the number of $\phi$ particles available for decay within the newly formed broken-phase volume is approximated by
\begin{equation}
dN_\phi=n_\phi^{\rm eq}\,d\mathcal{V}_{\rm B},
\end{equation}
where $n_\phi^{\rm eq}$ is the equilibrium number density of the $\phi$ particle immediately outside the advancing bubble wall in the symmetric phase.
The quantity $d\mathcal{V}_{\rm B}/dt$, represents the rate at which the broken-phase volume grows during the first-order phase transition, is reduced to
\begin{equation}
\frac{1}{\mathcal{V}_{\text{tot}}}\frac{d\mathcal{V}_{\rm B}}{dt} = - H T \, \frac{d P_{\rm s}(T)}{dT},
\end{equation}
where $P_{\rm s}(T)$ represents the fraction of total volume converted into the broken phase, provided in Eq. \ref{eq:False vacuum fraction-4-Temperature dependence}.
Therefore, the above expression in \eqref{eq:GW energy density /time} acts as the source for the GW energy density injected per unit time in the energy interval $\left[E_e, E_e+dE_e \right]$.

To track the production of gravitons and the subsequent evolution of the injected GW energy density we solve the differential Boltzmann equation (BE), expressed as
\begin{widetext}
\begin{eqnarray}
    \frac{d}{dT}\left(\frac{d\rho_{\text{GW}}}{d\ln{E_{e}}}\right)-\frac{4}{T}\frac{d\rho_{\text{GW}}}{d\ln{E_{e}}} = 
    E_{e}n^{\text{eq}}_{\phi}\left(T \right)\frac{d P_{\rm s}\left(T \right)} {dT} \frac{1}{\Gamma_{\phi\rightarrow\bar{\psi}\psi}} \frac{d\Gamma_{\phi\rightarrow\bar{\psi}\psi h}}{d\ln{E_{e}}},
    \label{eq:BE for GW energy density}
\end{eqnarray}
\end{widetext}
the derivation of which is detailed in Appendix \ref{sec:appendix-A}. 
The BE in Eq.~(\ref{eq:BE for GW energy density}) is solved numerically
between the temperatures $T_n$ and $T_f$, corresponding to the onset of
bubble nucleation and the completion of the FOPT characterised by $\Gamma_d(T_n) = H(T_n)$, $P_{\rm s}\left(T_f\right)=0$ respectively. During this period, gravitons are continuously
produced through the decay of the particle $\phi$, thereby sourcing the GW
energy density.

To relate the emitted GW spectrum to present-day observables, the graviton
energy $E_e$ emitted at a temperature $T\in[T_n,T_f]$ is expressed in terms
of the observed energy $E_0$ (or equivalently the observed frequency
$f_0$) as
\begin{equation}
E_e(E_0,T)
=
\frac{E_0\,T}{T_0}
\left(
\frac{g_{*s}(T)}
     {g_{*s}(T_0)}
\right)^{1/3}
=
\frac{2\pi f_0\,T}{T_0}
\left(
\frac{g_{*s}(T)}
     {g_{*s}(T_0)}
\right)^{1/3},
\end{equation}
where $T_0 = 6.626\times10^{-13}\,{\rm GeV}$ is the present temperature of
the Universe and $g_{*s}(T)$ denotes the effective number of entropy
degrees of freedom at temperature $T$. Substituting
$E_e(E_0,T)$ into Eq.~(\ref{eq:BE for GW energy density}) and solving the
equation over the interval $[T_n,T_f]$, one obtains the quantity
\[
\left.
\frac{d\rho_{\rm GW}}
     {d\ln f_0}
\right|_{T=T_f},
\]
which represents the GW energy density spectrum at the completion of the
phase transition, expressed as a function of the frequency $f_0$ observed
today.

After the completion of the phase transition at $T_f$, graviton production
ceases and the GW spectrum evolves solely through the expansion of the
Universe. Since gravitational waves behave as radiation, their energy
density redshifts as $a^{-4}$. Consequently, the present-day GW abundance
is obtained by redshifting the spectrum evaluated at $T=T_f$,
\begin{eqnarray}
\mathcal{h}^2 \Omega_{\rm GW}^{\rm brem} 
=
\frac{\mathcal{h}^2}{\rho_{{\rm cr},0}}
\left(
\frac{\xi T_0}{T_f}
\right)^4
\left.
\frac{d\rho_{\rm GW}}
     {d\ln f_0}
\right|_{T=T_f},
\label{eq:Total GW energy density today}
\end{eqnarray}
where, the quanity $\xi$, accounts for the change in the effective entropy degrees of freedom between the emission epoch and today, is defined as 
\begin{equation}
\xi=
\left(
\frac{g_{*s}(T_0)}
     {g_{*s}(T_f)}
\right)^{1/3}.
\end{equation}
Here, $\rho_{{\rm cr},0} = 8.1 \times 10^{-41} \mathcal{h}^2$ denotes the present critical energy density of the Universe, and $h$ is the reduced Hubble parameter.

\section{Results}
\begin{figure}[!htb]
	\centering
	\includegraphics[width=\linewidth]{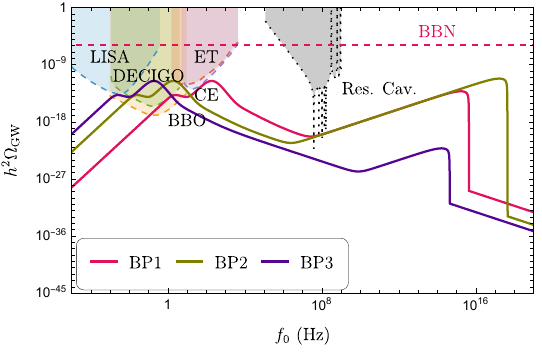}
	\caption{GW Spectrum for three benchmark points BP1, BP2 and BP3 as listed in Table~\ref{tab:Table for Benchmark points}.}
	\label{fig:GW-Spectra-BP12}
\end{figure}

Based on the above discussion, we note that all GW signals considered here are ultimately sourced by the first-order phase transition. 
While the conventional contributions arise from the macroscopic dynamics of bubble collisions, sound waves, and MHD turbulence, the same bubble nucleation process also activates a microscopic GW source through graviton bremsstrahlung from the decay of the scalar field $\phi$ within the expanding true-vacuum bubbles. 
As a consequence, the GW spectrum shown in Fig.~\ref{fig:GW-Spectra-BP12} exhibits multiple peaks (or bumps), each corresponding to a distinct GW production mechanism and therefore characterized by a different spectral shape.

The first peak, located at the lowest frequency in the form of a bump, arises from bubble
collisions. 
For the bubble collision contribution, the spectral shape function is given by~\cite{Jinno:2016vai,Athron:2023xlk}
\begin{equation}
S_{\rm col}(f)=
\frac{3.8\left(f/f_{\rm col}^{\rm peak}\right)^{2.8}}
     {1+2.8\left(f/f_{\rm col}^{\rm peak}\right)^{3.8}}.
\end{equation}
In the regime $f\ll f_{\rm col}^{\rm peak}$, the
denominator approaches unity and the spectral function behaves as
\begin{equation}
S_{\rm col}(f) \propto f^{2.8},
\end{equation}
resulting the GW spectrum to rise approximately as $\mathcal{h}^2 \Omega_{\rm GW}^{\rm col} \propto f^{2.8}$ toward the peak frequency. 
The spectrum reaches its maximum at the characteristic frequency $f_{\rm col}^{\rm peak}$ where $S_{\rm col}(f_{\rm col}^{\rm peak}) = 1$ takes the maximum value.
For frequencies above $f_{\rm col}^{\rm peak}$, the spectral function behaves as
\begin{equation}
S_{\rm col}(f) \propto f^{-1}.
\end{equation}
Hence, the bubble-collision GW contribution exhibits a relatively slow power-law fall-off at frequencies above the peak frequency.

The second peak originates from long-lived sound waves generated in the
plasma after bubble collisions. The associated spectral function~\cite{Caprini:2015zlo}
\begin{equation}
S_{\rm sw}(f)=
\left(\frac{f}{f_{\rm sw}^{\rm peak}}\right)^3
\left[\frac{7}{4+3\left(f/f_{\rm sw}^{\rm peak}\right)^2}\right]^{7/2}
\end{equation}
predicts that the spectrum rises as
\begin{equation}
\Omega_{\rm GW}^{\rm sw}(f)\propto f^3,
\end{equation}
for $f\ll f_{\rm sw}^{\rm peak}$. The GW spectrum strength is maximal around
$f_{\rm sw}^{\rm peak}$ as $S_{\rm sw}(f_{\rm sw}^{\rm peak}) = 1$ becomes maximum, after which the spectrum falls much more
rapidly, as
\begin{equation}
S_{\rm sw}(f)\propto f^{-4},
\end{equation}
for frequencies higher than $f_{\rm sw}^{\rm peak}$.
This steeper high-frequency suppression compared to the bubble-collision
signal is responsible for the initial decrease of the GW spectrum beyond
the sound-wave peak. However, as evident from
Fig.~\ref{fig:GW-Spectra-BP12}, the spectrum does not continue to follow the
same falling behavior over the entire frequency range. At higher
frequencies, the slope changes due to the contribution from
MHD turbulence generated in the plasma during the
phase transition.

The turbulent motion of the plasma acts as an additional long-lasting
source of GW and contributes predominantly at
frequencies above the sound-wave peak. As a result, when the sound-wave
signal begins to decrease as
\begin{equation}
\Omega_{\rm GW}^{\rm sw}(f)\propto f^{-4}, \label{eq:GW Sound wave falling behavior}
\end{equation}
the MHD turbulence component becomes increasingly important and modifies
the overall spectral shape. The GW spectrum in this frequency range is therefore determined by the superposition of the rapidly decreasing sound-wave contribution, characterized by Eq.~(\ref{eq:GW Sound wave falling behavior}), and the comparatively less suppressed turbulence contribution, which approaches the asymptote $\Omega_{\rm GW}^{\rm turb}\propto f^{-8/3}$ at sufficiently high frequencies. Consequently, the effective spectral slope of the total GW spectrum becomes less steep than that of the sound-wave contribution alone. This results in a change in the effective spectral shape of the total GW spectrum, which is visible in Fig.~\ref{fig:GW-Spectra-BP12} as an intermediate transition region from the sound-wave dominated regime to the MHD-turbulence dominated regime.

The third peak in the GW spectrum originates from graviton
bremsstrahlung accompanying the decay of the $\phi$ field. The
corresponding GW contribution has been obtained numerically by solving
Eq.~(\ref{eq:Total GW energy density today}). To gain analytical insight
into the frequency dependence of this contribution, we derive an
approximate expression for the GW spectrum from the numerical solution
of Eq.~(\ref{eq:Total GW energy density today}). The detailed derivation
is presented in Appendix~\ref{sec:appendix-B}. The
resulting analytical expression for the GW amplitude from graviton
bremsstrahlung is
\begin{align}
	\mathcal{h}^2\Omega_{\rm GW}^{\rm brem}
	\approx
	\left(
	\frac{m_\phi}{M_p}
	\right)^2
	\frac{f_0}{1.9\times 10^{17}~\rm Hz}\,
	\Theta\!\left(f_{\rm brem}^{\rm peak}-f_0\right),
	\label{eq:Approximate GW from bremsstrahlung}
\end{align}
where $\Theta(x)$ denotes the Heaviside step function.
Equation~(\ref{eq:Approximate GW from bremsstrahlung}) shows that the GW
spectrum generated from graviton bremsstrahlung increases linearly with
the observed frequency, $f_0$, up to a characteristic cut-off frequency
$f_{\rm brem}^{\rm peak}$. Beyond this frequency, graviton emission is
kinematically forbidden, leading to an abrupt termination of the
spectrum. The maximum observable graviton frequency is determined by the
kinematics of the $\phi$-particle decay and is given by
\begin{align}
	f_{\rm brem}^{\rm peak}
	=
	\frac{\xi T_0}{T_f}
	\frac{m_\phi/2}{2\pi},
	\label{eq: Cut-off frequency of graviton bremmss}
\end{align}
where $T_f$ denotes the temperature at the completion of the
FOPT and $\xi$ accounts for the redshifting of the emitted graviton frequency from the production epoch to the present Universe.

Therefore, unlike the GW spectra generated by bubble collisions, sound waves, and MHD turbulence, which exhibit a gradual power-law suppression beyond their respective peak frequencies, the
graviton-bremsstrahlung contribution displays a sharp cut-off at $f_{\rm brem}^{\rm peak}$ as a direct consequence of the decay kinematics.
An interesting feature of the FOPT-induced graviton-bremsstrahlung scenario is that the resulting GW amplitude is independent of the coupling $y_\phi$ governing the interaction between the $\phi$ field and the graviton ($h$). This can be seen explicitly from Eq.~(\ref{eq:Approximate GW from bremsstrahlung}), where the GW spectrum depends only on the ratio $m_\phi/M_p$ and the observed frequency $f_0$, with no explicit dependence on $y_\phi$. As demonstrated in Appendix~\ref{sec:appendix-A}, this independence results from the cancellation of $y_\phi$ between the graviton production rate and the lifetime of the $\phi$ particle. Consequently, the predicted GW signal is insensitive to the strength of the $\phi$--$\psi$ interaction, which constitutes a distinctive feature of our FOPT-embedded graviton-bremsstrahlung mechanism and stands in sharp contrast to conventional graviton-bremsstrahlung scenarios, where the GW amplitude generally depends explicitly on the interaction coupling.

Having established the characteristic spectral behaviour of the
graviton-bremsstrahlung contribution, we now investigate the interplay
between the conventional FOPT-generated GW sources and the
graviton-bremsstrahlung signal. 
To investigate the FOPT associated with the $\phi$ field, one must construct the corresponding finite-temperature effective potential, $V_{\rm eff}(\phi,T)$, and analyze its temperature evolution to track the formation of degenerate minima and the subsequent transition between them. 
Rather than adopting a specific form of $V_{\rm eff}(\phi,T)$, we perform our analysis in a model-agnostic framework, in which the dynamics of the FOPT are fully characterized by four macroscopic parameters: the percolation temperature $T_p$, the phase transition strength $\alpha$, the inverse duration parameter $\beta/H$, and the bubble wall velocity $v_w$. These quantities are treated as free parameters throughout our analysis.

The total GW spectrum is obtained by summing the contributions from the
three conventional FOPT sources together with the graviton
bremsstrahlung component, 
\begin{align}
	\mathcal{h}^2\Omega_{\rm GW}
	=
	\mathcal{h}^2\Omega_{\rm GW}^{\rm col}
	+
	\mathcal{h}^2\Omega_{\rm GW}^{\rm sw}
	+
	\mathcal{h}^2\Omega_{\rm GW}^{\rm turb}
	+
	\mathcal{h}^2\Omega_{\rm GW}^{\rm brem}.
\end{align}
The resulting multi-peak GW spectrum, together with the projected
sensitivities of future GW experiments such as LISA~\cite{LISA:2017pwj}, BBO~\cite{Yagi:2011wg,Crowder:2005nr,Corbin:2005ny,Harry:2006fi}, DECIGO~\cite{Yagi:2011wg,Kawamura:2006up}, ET~\cite{Punturo:2010zz,Hild:2010id,Sathyaprakash:2012jk,ET:2019dnz}, CE~\cite{LIGOScientific:2016wof,Reitze:2019iox}, Resonant Cavity experiment~\cite{Herman:2022fau}~\footnote{This is criticized as being overly optimistic in the literature \cite{Aggarwal:2025noe}}, is shown in
Fig.~\ref{fig:GW-Spectra-BP12}.

To illustrate the correlation between the FOPT dynamics and the GW
signal generated from graviton bremsstrahlung, we consider the three
benchmark points (BPs) listed in Table~\ref{tab:Table for Benchmark points}. In the first two benchmark points, all parameters are kept fixed except the percolation temperature $T_p$, which is chosen to be $10^7~{\rm GeV}$ and $10^5~{\rm GeV}$ for BP1 and BP2, respectively while the $\phi$ particle mass is chosen to $10^{13}$ GeV.
\begin{table}[!htb]
	\begin{tabular}{|c|c|c|c|c|c|}
		\hline
		& $m_\phi$ (GeV) & $\alpha$ & $\beta/H$ & $v_w$ & $T_p$ (GeV) \\
		\hline
		BP1 & $10^{13} $ & 0.1 & 50 & 0.5 & $10^7$  \\
		\hline
		BP2 & $10^{13} $ & 0.1 & 50 & 0.5 & $10^5$ \\
		\hline
        BP3 & $10^{9} $ & 0.1 & 50 & 0.5 & $10^4$ \\
        \hline
	\end{tabular}
	\caption{Benchmark points for GW spectrum calculation.}
	\label{tab:Table for Benchmark points}
\end{table}
Because the parameters $\alpha$, $\beta/H$, and $v_w$ are identical for
both benchmark points, the amplitudes of the FOPT-generated GW
components remain nearly unchanged. However, the characteristic peak
frequencies of the bubble collision, sound wave, and MHD turbulence
signals are shifted according to the different values of the
percolation temperature $T_p$, resulting in the displacement of the
first two peaks of the total GW spectrum.

A similar dependence is observed for the graviton-bremsstrahlung
contribution. Although the mass of the decaying particle $\phi$ is
chosen to be the same for both benchmark points BP1 and BP2, the cut-off frequency
of the bremsstrahlung spectrum,
$f_{\rm brem}^{\rm peak}$, scales linearly with
$m_\phi/T_p$ (as for same $\beta/H$, $T_f \propto T_p$), as evident from
Eq.~(\ref{eq: Cut-off frequency of graviton bremmss}). Consequently,
decreasing the percolation temperature shifts the
graviton-bremsstrahlung peak toward higher frequencies. Furthermore,
since the GW amplitude below the cut-off frequency satisfies
$\mathcal{h}^2\Omega_{\rm GW}^{\rm brem}\propto f_0$
[cf. Eq.~(\ref{eq:Approximate GW from bremsstrahlung})], the change in
$f_{\rm brem}^{\rm peak}$ also modifies the peak amplitude of the
bremsstrahlung signal. Therefore, despite the microscopic parameters of
the decay process remaining unchanged, the bremsstrahlung-induced GW
spectrum is directly correlated with the macroscopic dynamics of the
phase transition through the percolation temperature. This establishes
a direct connection between the high-frequency GW signal from graviton
bremsstrahlung and the underlying FOPT dynamics.

\begin{figure}[!h]
	\centering
	\includegraphics[width=\linewidth]{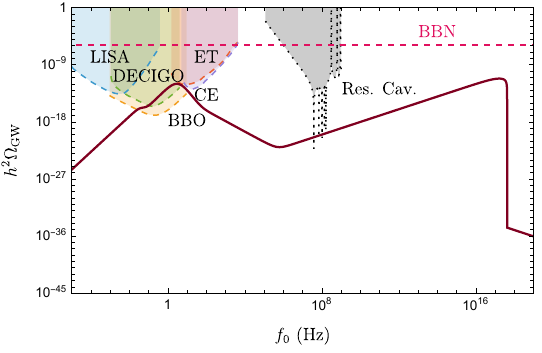}
	
	\caption{Multi-peaked stochastic GW spectrum for the benchmark point in Table~\ref{tab:Table for Benchmark point 4}, illustrating the complementary low- and high-frequency detection prospects in future GW experiments.}
	\label{fig:GW-Spectra-BP4}
\end{figure}

Finally, we highlight a distinctive observational signature of our
FOPT-induced graviton-bremsstrahlung scenario. As shown in
Fig.~\ref{fig:GW-Spectra-BP4}, the conventional GW signal generated by
the FOPT can be probed in the low-frequency
window by future space- and ground-based GW observatories such as BBO,
DECIGO, ET, and CE. Remarkably, the same phase transition simultaneously
produces a high-frequency GW signal through graviton bremsstrahlung,
which lies within the sensitivity range of Resonant Cavity experiment.
Therefore, a simultaneous observation of the low-frequency FOPT-induced
GW spectrum together with its high-frequency graviton-bremsstrahlung
counterpart would constitute a unique and robust signature of our
scenario. The corresponding benchmark point BP3 is shown in Table~\ref{tab:Table for Benchmark point 4}. 
\begin{table}[!htb]
	\begin{tabular}{|c|c|c|c|c|c|}
		\hline
		& $m_\phi$ (GeV) & $\alpha$ & $\beta/H$ & $v_w$ & $T_p$ (GeV) \\
		\hline
		BP4 & $10^{13} $ & 0.03 & 100 & 0.65 & $10^5$  \\
		\hline
	\end{tabular}
  \caption{Benchmark point adopted for the GW spectrum presented in Fig~\ref{fig:GW-Spectra-BP4}.}
  \label{tab:Table for Benchmark point 4}
\end{table}
Such a multi-band GW detection would not only provide
compelling evidence for the occurrence of a first-order phase
transition, but would also establish a direct connection between the
macroscopic dynamics of the phase transition and the microscopic
graviton-bremsstrahlung process, thereby offering a distinctive avenue
to test the proposed framework.

\section{Conclusion}
In this work, we have proposed a new mechanism for generating a multi-peaked stochastic GW background, in which all spectral features originate from a \textit{single} cosmological first-order phase transition. In addition to the conventional GW sources arising from FOPT dynamics, we have identified a previously unexplored microscopic GW source generated by graviton bremsstrahlung during the decay of the scalar field responsible for the same phase transition. Consequently, the predicted GW spectrum simultaneously encodes information about the macroscopic dynamics of bubble expansion and plasma evolution, as well as the microscopic particle interactions associated with the phase-transition scalar.

A particularly distinctive aspect of our framework is that the
high-frequency graviton-bremsstrahlung signal is not an independent
cosmological relic, but is intrinsically correlated with the dynamics of the underlying phase transition. This is in sharp contrast to previously proposed multi-peaked GW scenarios, where different spectral features typically originate from multiple cosmological epochs or from subsequent unrelated sources such as domain walls, cosmic strings, primordial black holes, or multiple successive phase transitions. In our scenario, all peaks are generated during the same FOPT through different physical mechanisms, providing a unified origin for the entire GW spectrum. Furthermore, since graviton bremsstrahlung follows solely from the universal coupling of gravity to the energy-momentum tensor, the mechanism is generic and can be incorporated into any FOPT framework, without relying on the details of a particular scalar potential model.

The graviton-bremsstrahlung contribution itself possesses several
remarkable properties. We have shown that its spectrum increases
approximately linearly with frequency before terminating abruptly at a
kinematically determined cut-off frequency, in striking contrast to the gradual power-law suppression exhibited by the conventional
FOPT-generated GW sources. We have further demonstrated that the GW
amplitude is independent of the Yukawa coupling governing the scalar
decay, owing to an exact cancellation between the coupling dependence of the graviton production rate and the lifetime of the decaying scalar. Moreover, the cut-off frequency is directly determined by the
percolation temperature, thereby establishing a direct link between the microscopic graviton-bremsstrahlung signal and the macroscopic dynamics of the phase transition. 
our results demonstrate that high-frequency gravitational
waves need not constitute an independent cosmological background, but
can instead serve as a correlated microscopic probe of the same
first-order phase transition responsible for the conventional
stochastic GW signal. This establishes graviton bremsstrahlung as a new observational window connecting cosmological phase transitions with microscopic particle interactions.

Finally, our benchmark analysis demonstrates that the conventional
FOPT-generated GW signal is potentially detectable in the low-frequency band by future interferometers such as BBO, DECIGO, ET, and CE, while the correlated graviton-bremsstrahlung signal can simultaneously appear in the high-frequency window accessible to Resonant Cavity experiment.
A simultaneous observation of these low- and high-frequency GW signals
would not only constitute compelling evidence for a cosmological
FOPT, but would also provide a direct probe of
both the macroscopic phase-transition dynamics and the microscopic
particle interactions responsible for graviton bremsstrahlung. We
therefore expect the mechanism proposed in this work to open a new
avenue for exploring particle physics beyond the Standard Model through future multi-band gravitational-wave observations.

\appendix

\section{Derivation of the Boltzmann Equation}
\label{sec:appendix-A}
We begin with the 1-particle density function (1-PDF) $f$ associated to the graviton energy density per energy bin $[E,E+dE]$ as given by
\begin{equation}
f=\frac{\left(2\pi\right)^{3}dN}{\left(d^{3}p\right)\left(dV\right)}=\frac{2\pi^{2}}{p^{2}E}\frac{Edn}{dp}=\frac{2\pi^{2}}{p^{2}E}\frac{dE}{dp}\frac{d\rho}{dE}, \label{eq:1-pdf-def}
\end{equation}
where $N$ and $n$ denote number of gravitons and density of gravitons respectively. All other symbols have their usual meaning. Using the relation $E^{2}=p^{2}+m^{2}$, we have $\frac{dE}{dp}=\frac{p}{E}$.
Additionally, for gravitons, $m=0$ and $E=p$, thus we have
\begin{equation}
f=\frac{2\pi^{2}}{pE^{2}}\frac{d\rho}{dE}=\frac{2\pi^{2}}{E^{3}}\frac{d\rho}{dE}.\label{eq:1-pdf-eq-1}
\end{equation}
Now the Liouville theorem gives us the corresponding Boltzmann equation
\begin{equation}
\left.\frac{df}{dt}\right|_{\text{streamline}}=\left.\frac{\partial f}{\partial t}\right|_{\text{coll}},\label{eq:1-pdf-bte}
\end{equation}
which states that the total change in the 1-PDF along the trajectory
of a single graviton (expressed by the streamline derivative above) in the expanding FRW universe is given by the
change of 1-PDF only due to collision (here production from $\phi\rightarrow\bar{\psi}\psi h$
decay, expressed by the collisional derivative).

To evaluate the streamline derivative on the LHS of Eq. \ref{eq:1-pdf-bte}, we substitute \eqref{eq:1-pdf-eq-1} in it, 
which results to 
\begin{equation}
\frac{df}{dt}=\frac{d}{dt}\left[\frac{2\pi^{2}}{E^{3}}\frac{d\rho}{dE}\right]=\frac{2\pi^{2}}{E^{3}}\left[\frac{d}{dt}\frac{d\rho}{dE}-\frac{3}{E}\frac{dE}{dt}\frac{d\rho}{dE}\right].\label{eq:bte-lhs}
\end{equation}
In the expanding FRW universe, $E$ redshifts as $E\propto a^{-1}$
and $\frac{d\ln{E}}{d\ln{a}}=-1$. Thus, we have
\begin{equation}
\frac{1}{E}\frac{dE}{dt}=H\frac{a}{E}\frac{dE}{da}=H\frac{d\ln{E}}{d\ln{a}}=-H.\label{eq:redshift-rate}
\end{equation}
Substituting \eqref{eq:redshift-rate} into \eqref{eq:bte-lhs}, we obtain
\begin{equation}
\frac{df}{dt}=\frac{2\pi^{2}}{E^{3}}\left[\frac{d}{dt}\frac{d\rho}{dE}+3H\frac{d\rho}{dE}\right]=\left.\frac{\partial f}{\partial t}\right|_{\text{coll}}.\label{eq:bte-1}
\end{equation}
Thus, the Boltzmann equation in \eqref{eq:1-pdf-bte} takes the following form 
\begin{equation}
\frac{d}{dt}\left(\frac{d\rho}{dE}\right)+3H\frac{d\rho}{dE}=\frac{E^{3}}{2\pi^{2}}\left.\frac{\partial f}{\partial t}\right|_{\text{coll}}=\left.\frac{\partial}{\partial t}\left(\frac{d\rho}{dE}\right)\right|_{\text{coll}}.\label{eq:bte-2}
\end{equation}

When a $\phi$ particle decays, the probability that it will produce
a graviton in the energy bin $\left[E,E+dE\right]$ is given by the
branching ratio $\frac{1}{\Gamma_{\phi\rightarrow\bar{\psi}\psi}}\frac{d\Gamma_{\phi\rightarrow\bar{\psi}\psi h}}{dE}$.
The rate of change of $\phi$ particle density due to the expansion
of the bubble of broken phase is given by $n^{\text{eq}}_{\phi}\left(T\right)\frac{dP_{\rm s}}{dt}$.
Thus, $\frac{d\rho}{dE}$ changes at the rate
\begin{equation}
\left.\frac{\partial}{\partial t}\left(\frac{d\rho}{dE}\right)\right|_{\text{coll}}=En^{\text{eq}}_{\phi}\left(T\right)\frac{dP_{\rm s}}{dt}\frac{1}{\Gamma_{\phi\rightarrow\bar{\psi}\psi}}\frac{d\Gamma_{\phi\rightarrow\bar{\psi}\psi h}}{dE},\label{eq:bte-3}
\end{equation}
purely due to $\phi$ decay. Substituting it back into \eqref{eq:bte-2},
we obtain the Boltzmann equation as given by 
\begin{equation}
\frac{d}{dt}\left(\frac{d\rho}{dE}\right)+3H\frac{d\rho}{dE}=En^{\text{eq}}_{\phi}\left(T\right)\frac{dP_{\rm s}}{dt}\frac{1}{\Gamma^{\left(2\right)}}\frac{d\Gamma^{\left(3\right)}}{dE}.\label{eq:bte-4}
\end{equation}
Since the GW spectrum is related to $\frac{d\rho}{d\ln{E}}$ via 
\begin{equation}
\mathcal{h}^2\Omega_{\text{GW}}^{brem}=\frac{\mathcal{h}^2}{\rho_{\text{cr}}}\left(\frac{\xi T_{0}}{T_{f}}\right)^{4}\frac{d\rho}{d\ln{E}},\label{eq:OmegaGW-appen}
\end{equation}
it is useful to rewrite the Boltzmann equation using evolution of $\frac{d\rho}{d\ln{E}}=E\frac{d\rho}{dE}$
instead. Now the first term in the LHS of \eqref{eq:OmegaGW-appen} can be written as 
\begin{align}
\frac{d}{dt}\left(\frac{d\rho}{dE}\right)=\frac{d}{dt}\left(\frac{1}{E}\frac{d\rho}{d\ln{E}}\right)=\nonumber \\
\frac{1}{E}\left[\frac{d}{dt}\left(\frac{d\rho}{d\ln{E}}\right)-\frac{1}{E}\frac{dE}{dt}\frac{d\rho}{d\ln{E}}\right], 
\end{align}
which along with \eqref{eq:redshift-rate} becomes 
\begin{equation}
\frac{d}{dt}\left(\frac{d\rho}{dE}\right)=\frac{1}{E}\left[\frac{d}{dt}\left(\frac{d\rho}{d\ln{E}}\right)-H\frac{d\rho}{d\ln{E}}\right].
\end{equation}
Substituting back into \eqref{eq:bte-4}, we obtain the Boltzmann equation
\begin{equation}
\frac{d}{dt}\left(\frac{d\rho}{d\ln{E}}\right)+4H\frac{d\rho}{d\ln{E}}=En^{\text{eq}}_{\phi}\left(T\right)\frac{dP_{\rm s}}{dt}\frac{1}{\Gamma_{\phi\rightarrow\bar{\psi}\psi}}\frac{d\Gamma_{\phi\rightarrow\bar{\psi}\psi h}}{d\ln{E}}.
\end{equation}
This Boltzmann equation is easier to solve if we trade time $t$ in favor of
temperature $T$. Using
\begin{equation}
\frac{d}{dt}=-HT\frac{d}{dT},
\end{equation}
we obtain the following form of the Boltzmann equation
\begin{equation}
\frac{d}{dT}\left(\frac{d\rho}{d\ln{E}}\right)-\frac{4}{T}\frac{d\rho}{d\ln{E}}=En^{\text{eq}}_{\phi}\left(T\right)\frac{dP_{\rm s}}{dT}\frac{1}{\Gamma_{\phi\rightarrow\bar{\psi}\psi}}\frac{d\Gamma_{\phi\rightarrow\bar{\psi}\psi h}}{d\ln{E}},\label{eq:bte-final}
\end{equation}
which is used in the main text.

\section{Approximate Behavior of the GW Spectra}
\label{sec:appendix-B}
Note that \eqref{eq:bte-final} can be written as a total derivative as follows.
\begin{equation}
T^{4}\frac{d}{dT}\left(\frac{1}{T^{4}}\frac{d\rho}{d\ln{E}}\right)=En^{\text{eq}}_{\phi}\left(T\right)\frac{dP_{\rm s}}{dT}\frac{1}{\Gamma_{\phi\rightarrow\bar{\psi}\psi}}\frac{d\Gamma_{\phi\rightarrow\bar{\psi}\psi h}}{d\ln{E}}.\label{eq:bte-final-1}
\end{equation}
Here the number density of $\phi$ particles in the symmetric phase is given by 
\begin{equation}
n^{\text{eq}}_{\phi}\left(T\right)=\frac{\zeta\left(3\right)}{\pi^{2}}T^{3},\label{eq:nphieq}
\end{equation}
due to it massless nature there. Substituting \eqref{eq:nphieq} into
\eqref{eq:bte-final-1} and using the redshift, $E=2\pi f=\frac{2\pi T}{\xi T_{0}}f_{0}$,
we obtain 
\begin{equation}
\frac{d}{dP_{\rm s}}\left(\frac{1}{T^{4}}\frac{d\rho}{d\ln{E}}\right)=2\pi f_{0}\frac{1}{\xi T_{0}}\frac{\zeta\left(3\right)}{\pi^{2}}\frac{1}{\Gamma_{\phi\rightarrow\bar{\psi}\psi}}\frac{d\Gamma_{\phi\rightarrow\bar{\psi}\psi h}}{d\ln{E}}.\label{eq:bte-final-1-1}
\end{equation}
Here we have changed the independent variable from $T$ to $P_{\rm s}$. 

Now integrating
between $P_{\rm s}=0$ and $P_{\rm s}=1$, we obtain
\begin{equation}
\frac{d\rho}{d\ln{E}}=\frac{2\pi f_{0}}{\xi T_{0}}\frac{\zeta\left(3\right)}{\pi^{2}}T^{4}_{f}\int^{1}_{0}dP_{\rm s}\frac{1}{\Gamma_{\phi\rightarrow\bar{\psi}\psi}}\frac{d\Gamma_{\phi\rightarrow\bar{\psi}\psi h}}{d\ln{E}}.\label{eq:bte-final-1-1-1}
\end{equation}
Finally, using \eqref{eq:OmegaGW-appen}, we get the GW spectrum
\begin{equation}
\mathcal{h}^2\Omega_{\text{GW}}^{\rm brem}=\frac{\zeta\left(3\right)}{\pi^{2}}\frac{\mathcal{h}^2\left(\xi T_{0}\right)^{3}}{\rho_{\text{cr}}}\left(2\pi f_{0}\right)\int^{1}_{0}dP_{\rm s}\frac{1}{\Gamma_{\phi\rightarrow\bar{\psi}\psi}}\frac{d\Gamma_{\phi\rightarrow\bar{\psi}\psi h}}{d\ln{E}}.
\end{equation}
In the limit $x\rightarrow0$ and $y\rightarrow0$ (corresponding to $E\rightarrow 0$ and $m_{\psi}\rightarrow 0$), the \eqref{eq:2-body-decay-rate} and \eqref{eq:3-body-ddr} of the main text then become
\begin{equation}
\Gamma_{\phi\rightarrow\bar{\psi}\psi}=\frac{y_\phi^{2}}{8\pi}m_{\phi},
\end{equation}
and
\begin{equation}
\frac{d\Gamma_{\phi\rightarrow\bar{\psi}\psi h}}{d\ln{E}}=\frac{y_\phi^{2}}{64\pi^{3}}m_{\phi}\left(\frac{m_{\phi}}{M_{p}}\right)^{2},
\end{equation}
respectively. Thus, the branching ratio of the 3-body decay to 2-body decay becomes
\begin{equation}
\frac{1}{\Gamma_{\phi\rightarrow\bar{\psi}\psi}}\frac{d\Gamma_{\phi\rightarrow\bar{\psi}\psi h}}{d\ln{E}}=\frac{1}{8\pi^{2}}\left(\frac{m_{\phi}}{M_{p}}\right)^{2}.\label{eq:branching-ratio}
\end{equation}
Assuming an instantaneous FOPT and using \eqref{eq:OmegaGW-appen}, we
obtain the approximate spectrum 
\begin{equation}
\mathcal{h}^2\Omega_{\text{GW}}^{\rm brem}\approx\frac{\zeta\left(3\right)}{8\pi^{4}}\frac{\mathcal{h}^2}{\rho_{\text{cr}}}\left(\frac{m_{\phi}}{M_{p}}\right)^{2}\left(\xi T_{0}\right)^{3}\left(2\pi f_{0}\right)\Theta\left(f^{\text{peak}}_{\rm brem}-f_{0}\right),
\end{equation}
where the $\Theta$ function cuts off the spectrum at the maximum
frequency with which gravitons are ever emitted $f^{\text{peak}}_{\rm brem}=\frac{\xi T_{0}}{4\pi}\frac{m_{\phi}}{T_{f}}$.

\bibliography{references}

\end{document}